\documentclass[twocolumn,showpacs,superscriptaddress,preprintnumbers,amsmath,amssymb,aps,floatfix]{revtex4}

\usepackage{graphicx}
\usepackage{amsmath}
\usepackage{subfigure}

\begin{document}


\title{Statistical interaction description of Pauli crystals in two-dimensional systems
        of harmonically confined fermions}

\author{Orion Ciftja}
\email{ogciftja@pvamu.edu}
\affiliation{Department of Physics, Prairie View A\&M University,
                Prairie View, TX 77446, USA}

\author{Josep Batle}
\email{jbv276@uib.es}
\affiliation{Departament de F\'{\i}sica, Universitat de les Illes Balears, 
                07122 Palma de Mallorca, Balearic Islands, Spain}

\date{July 23, 2019}

\begin{abstract}

It has been conjectured that
the Pauli exclusion principle alone may be 
responsible for a particular geometric arrangement of confined systems of 
identical fermions even when there is no interaction between them.
These geometric structures, called Pauli crystals, are predicted for 
a two-dimensional system of free fermions under harmonic confinement.
It is assumed that the system consists of neutral fermionic atoms
with their spins frozen (spin-polarized) in order 
to avoid any form of electromagnetic interaction.
These crystalline patterns emerge as the most frequent configurations 
seen in a large collection of single-shot pictures of the system.
In this work, we pursue the possiblity of this outcome and
consider a theoretical model  that may capture both qualitatively and quantitatively 
key features of the above mentioned setup.
Our approach treats a quantum system of non-interacting fermions
as an effective classical system of particles that interact with an effective 
{\it statistical interaction} potential that mimics the quantum statistics.
For this model, we consider two-dimensional few-body systems of harmonically
confined particles that interact with a statistical potential 
and calculate analytically the minimum energy 
configuration for specific values of relevant parameters.
The results for $N=3$ and $6$ particles show that the minimum energy 
configuration corresponds to and is in good quantitative 
agreement with the reported values of Pauli crystals seen in 
single-shot imaging data obtained via the configuration density technique.
Numerical results for larger systems of $N=15$ and $30$ particles
show that the crystalline configurations observed are not  the same
as the classical Wigner crystal structures that emerge
should the confined charged particles interact with a Coulomb potential.
An important question floated is whether such crystalline structures do really exist in a 
quantum system or whether they are artifacts of the methods used to analyze them.

%
%
%


 
\end{abstract}


\maketitle


\section{Introduction}
\label{sec-intro}

Pauli's exclusion principle~\cite{pauli} enforces high-order correlations 
in systems of identical fermions and 
its consequences are well known in quantum physics~\cite{dirac,contemporary,cohen}.
This principle applies to fermions (such as electrons) and not bosons~\cite{2014c}. 
Pauli's exclusion principle basically states that no two fermions can occupy 
the same quantum state and, thus,
anti-symmetrization of the full wave function is required~\cite{slater,2015f}.
Recent work~\cite{rakshit,gajda1,gajda2} focused on confined systems of free fermions 
in a harmonic trap argues that quantum correlations imposed by Pauli's exclusion principle
may lead to the stabilization of crystalline structures, called {\it Pauli crystals}.
These crystalline structures emerge even when there is 
no mutual interaction between fermions.
It was argued that a crystal-like structure is observed 
when many identical fermions at very low temperature 
are confined within an external two-dimensional (2D) harmonic trap.
The authors of this work~\cite{rakshit,gajda1,gajda2} 
showed how to extract these geometric structures 
from multiple single-shot pictures of the many-body system
using a numerical approach known as the configuration density technique. 
%
%
%
The authors of Ref.[\citenum{rakshit,gajda1,gajda2}] refer to the results
obtained by using the numerical configuration density approach 
as "single-shot imaging experiments".
Although we will refer to these works with a similar terminology, 
it is our understanding that the works that we just mentioned 
are theoretical (they represent computational processing of large data sets).
As far as we know, any experimental verification of the existence
of the Pauli crystals does not exist at the moment. 

It is important to remark that Pauli crystals are different from 
Wigner crystals of electrons~\cite{Wig} 
or Coulomb crystals of ions since there is no interaction between fermions
in the case of Pauli crystals.  
On the other hand, it is known that Wigner crystals of electrons or 
Coulomb crystals of ions are stabilized by interaction effects 
between particles~\cite{Grimes}.
This means that Pauli crystals have a very different origin. 
Absence of any interaction between fermions is a pre-requisite  
to observe the effects of quantum statistics. Thus, one may detect  these
structures in few-body systems of non-interacting charge-neutral spin-polarized 
atomic fermions under 2D harmonic confinement at
ultracold temperatures~\cite{rakshit,gajda1,gajda2}.


In this work we introduce a theoretical model for a Pauli crystal
for the setup considered in Ref.[\citenum{rakshit,gajda1,gajda2}].
The model allows us to treat a quantum system of free fermions 
as a classical ensemble of particles interacting with
a {\it statistical interaction} potential~\cite{uhlenbeck,aop}.
The {\it statistical interaction} potential mimics the quantum statistics
of the particles.
%
We consider small systems of $N=3$ and $N=6$ particles confined in a 2D harmonic trap 
and calculate exactly the minimum energy configuration corresponding 
to various parameters (such as temperature, etc.).
%
%
The results obtained for systems with $N=3$ and $N=6$ particles
show that the minimum energy configurations in this model 
correspond to the Pauli crystal configurations reported 
in Ref.[\citenum{rakshit,gajda1,gajda2}]
and are in good quantitative agreeement with the 
equilibrium parameters reported in Ref.[\citenum{rakshit,gajda1,gajda2}].
Numerical results for larger systems with $N=15$ and $N=30$ particles
show that the crystalline configurations observed are not  the same
as the corresponding classical Wigner crystal structures of 
confined charges in a 2D harmonic trap.
%
%
%


The article is composed as follows. 
In Section~\ref{sec-exp} we briefly describe the methods used in 
single-shot imaging experiments.
In Section~\ref{sec-model} we introduce the theory and model for the case of 
particles under 2D harmonic confinement.
Section~\ref{sec-fewbody} contains the key results that apply to few-body systems 
of particles. 
A brief discussion and some concluding remarks are found in Section~\ref{sec-discussion}.


\section{Single-shot imaging experiments of free fermions in a harmonic trap}
\label{sec-exp}

In this section we provide a brief description of recent single-shot imaging
experiments which seem to provide hints for the existence of Pauli crystals.
The discussions ate based and rely on the work described in 
Ref.[\citenum{rakshit,gajda1,gajda2}].
The system considered is a cloud of non-interacting fermions confined 
in a 2D isotropic harmonic trap with frequency, $\omega$.
The fermions have their spins frozen (the system is spin-polarized).
For such a case, the quantum one-particle wave functions would be written as:
 \begin{equation} 
 \phi_{n_x n_y}(x,y)=N_{n_x n_y} \,
 \exp\Bigl(-\frac{x^2+y^2}{2 \, l_0^2} \Bigr) \, H_{n_x} \Bigl(\frac{x}{l_0}\Bigr)
                                                \, H_{n_y}\Bigl( \frac{y}{l_0} \Bigr) \ ,
 \label{wf}
 \end{equation}
where
$N_{n_x n_y}$ is the appropriate normalization constant,
$H_{n}(x)$ is a Hermite polynomial,
$n_{x,y}=0, 1, \ldots$ are the allowed quantum numbers
and
 \begin{equation} 
 l_0=\sqrt{\frac{\hbar}{m \, \omega}} \ ,
 \label{oscillatorlength}
 \end{equation}
is the so-called harmonic oscillator length.
The number of quantum states with energy,
\begin{equation} \label{energy-quantum}
E_{n_x n_y}=\hbar \, \omega \, (n_x+n_y+1) \ ,
\end{equation}
%
is $(n_x+n_y+1)$ and this represents the degeneracy of that energy value.
An energy shell is said to be filled 
if all quantum states corresponding to that energy level are occupied.
Systems with $N=1, 3, 6, 10, 15, 21, \ldots $ particles correspond to filled shells.
%
%
A Slater determinant wave function of occupied one-particle states,
\begin{equation} 
\Psi(\vec{r}_1, \ldots, \vec{r}_N)=\frac{1}{\sqrt{N!}} 
                                              Det \Bigl\{ \phi_{\vec{n}_i}(\vec{r}_j)  \Bigr\}  \ ,
\label{slater}
\end{equation}
%
%
represents 
the many-body ground state wave function consistent with Pauli's exclusion principle.
In the above notation $\vec{n}_i=(n_{ix}, n_{iy})$ and $\vec{r}_{j}$ is 
a 2D position vector for each of the $j=1, \ldots , N$ particles.

In the single-shot measurements reported in Ref.[\citenum{rakshit,gajda1,gajda2}], 
one attempts to determine the position configuration of $N$ fermions 
that maximizes the value of the probability distribution of the system, 
namely the configuration that maximizes the value of the modulus squared 
of the wave function,  $ |\Psi(\vec{r}_1, \ldots, \vec{r}_N) |^2$.
A single-shot picture of the system allows one to obtain the particles's positions.
Since positions of particles in a quantum system are probabilistic variables
one searches for the most probable ones.  
To this effect, the above-mentioned authors have found efficient ways 
to process a very large number of single-shots using a method known
as the configuration density technique.
A single-shot measurement leads to the corresponding sets
of particle's position configurations. 
The large amount of data for a multitude of single-shots 
are analyzed and from there one can extract the most probable 
configurations of the fermions.
In a nutshell, after processing a multitude of single-shot imaging measurements,
the results seem to suggest that the most probable configurations observed 
have a distinct crystalline nature and are universal if the number $N$ of fermions 
in a 2D harmonic trap corresponds to filled shells~\cite{gajda1}.
Specifically speaking, an equilateral triangle for $N=3$ and a pentagon (with an additional
atom at the center) for $N=6$ fermions is observed.

The above-mentioned crystalline patterns are extracted from the measured 
noisy structure by using image processing techniques~\cite{gajda2}.
An instantaneous picture of the atoms gives a set of $N$ positions vectors.
However, as mentioned earlier, 
a single picture of the many-body system in its ground state cannot reveal
anything about the underlying configuration predicted by using the probability
distribution~\cite{rakshit}.
In order to circumvent  this problem and obtain the $N$-particle correlation function
a multi-shot analysis procedure is employed in order
to uncover the structure of the most probable configuration.
The details of how to analyze the outcomes of single-shot measurements 
is explained in Ref.[~\citenum{rakshit,gajda1,gajda2}].
The number of single-shot pictures that leads to identifiable structures
for the case of $N=3$ and $N=6$ fermions can be as small as $10^3$ but
preferably should be as large as possible~\cite{rakshit}. 
The results reported in Ref.[~\citenum{gajda1,gajda2}] 
for a system of $N=6$ fermions are obtained 
after image processing is applied to an ensemble of  $2 \times 10^6$ configurations
at a temperature, $T= 1 \, \hbar \omega/k_B$.
The crystalline structure in this case has one particle at the center of the 
harmonic trap  and five particles in an outer pentagon structure each at a distance 
$r/l_0=1.265$ from the center.

\begin{figure}[!thbp]
\begin{center}
\includegraphics[width=9cm]{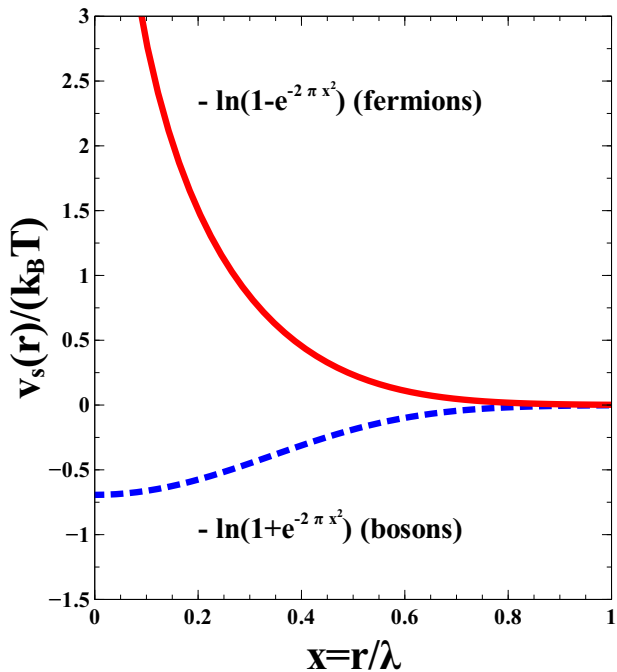}
\caption{
            Dependence of statistical interaction potential $v_s(r)/(k_B \, T)$ 
            as a function of $x=r/\lambda$ for fermions (top) and bosons (bottom). 
            The value of the potential decays fast and is very close to zero for
           $r \approx \lambda$. }
\label{fig1}
\end{center}
\end{figure}
%
%

%
   \section{Model}
\label{sec-model}

In quantum statistical mechanics,
the mean thermal wavelength parameter, $\lambda$
is defined as:
%
%
\begin{equation} \label{lambdaT}
\lambda = \sqrt{\frac{2 \, \pi \, \hbar^2}{m \, k_B \, T}} \ ,
\end{equation}
where
$m$  represents the mass of the particle,
$k_B$ is Boltzmann's constant
$T$ is the absolute temperature of the system in Kelvin degrees
and
$\hbar$ is the reduced Planck's constant.
The ratio of the mean thermal wavelength to the mean interparticle
distance can be related to the "indistinguishability" of quantum particles.
If $\lambda$ is smaller than typical interparticle separations, the system
of fermions will be nondegenerate and will be approximately classical satisfying 
Boltzmann statistics.
This quantity is called the thermal wavelength because has the same order of
magnitude as the de Broglie wavelength of a particle of mass $m$
with energy $k_B T$.
The statistical correlations between quantum particles are expected to be
irrelevant if temperature is so high that $\lambda$ 
is much smaller than the average interparticle distance.
For such conditions, the system can be treated as an ideal gas.
The first quantum correction to the classical partition function of an ideal gas
can be rigorously calculated by following a recipe developed by
Uhlenbeck et al.~\cite{uhlenbeck}.
The central notion of the method is the mapping of non-interacting fermions
into a classical interacting system via the so-called Uhlenbeck's 
effective temperature-dependent {\it statistical interaction} potential.
This correction has the same effect as endowing the particles with an inter-particle
effective statistical interaction potential, $v_s(r)$ 
and treating the system classically.
The treatment is general and applies to both fermions and bosons.
The statistical potential between quantum particles (fermions or bosons) that arises 
from the symmetry properties of the $N$-particle wave function can be written as:
\begin{equation} \label{statpot}
v_s(r) = -k_B \, T \, \ln \Bigl[ 1 \pm \exp\Bigl( -2 \, \pi \, r^2/\lambda^2 \Bigr) \Bigr] \ ,
\end{equation}
%
%
where the "$\pm$" sign applies to, respectively,  bosons/fermions. 
As seen from Fig.~\ref{fig1}, 
the interaction of two bosons is effectively "attractive"  
while that of two of fermions is highly "repulsive" at short separation distances.
In few words, the effective interaction mimics Pauli-repulsion effects for the case
of fermions.
%
%
%
Reduction of the temperature would lead to an increase of the
range of the statistical potential.
This implies that very low temperatures are desirable for any possible
experimental observation of such purely statistical quantum effects. 
%
%


The current model inspired by this approach consists of 
$N$ interacting classical particles under an isotropic 2D harmonic confinement:
%
%
\begin{equation} \label{emodel}
E=\sum_{i<j}^{N}   v( r_{ij} )+ \frac{m}{2} \, \omega^2 \sum_{i=1}^{N} r_i^2 ,
\end{equation}
where 
$v(r_{ij})=-k_B \, T \, \ln \Bigl[ 1- \exp\Bigl( -2 \, \pi \, r_{ij}^2/\lambda^2 \Bigr) \Bigr]$
is the effective fermionic statistical interaction potential,
$r_{ij}=|\vec{r}_i-\vec{r}_j|$ represents the 2D inter-particle separation distance,
$m$ is the mass,
$\omega$ is the angular frequency of the harmonic potential
and
$\vec{r}_i$ are 2D position vectors.
The energy of the system can be conveniently expressed in units of $\hbar \, \omega$
and can be written as:
%
%
\begin{equation} 
\frac{E(\alpha)}{\hbar \, \omega}=
-\alpha \sum_{i<j}^{N} \ln \Bigl[ 1-\exp \Bigl( -\alpha \, \frac{r_{ij}^2}{l_0^2}  \Bigr) \Bigr]+
                      \frac{1}{2}\sum_{i=1}^{N} \frac{r_i^2}{l_0^2} \ ,
\label{Etotal}
\end{equation}
%
where
\begin{equation} 
\label{alpha}
\alpha=\frac{k_B \, T}{\hbar \, \omega}  \ ,
\end{equation}
is a dimensionless temperature parameter and
$l_0$ is the harmonic oscillator length defined in Eq.(\ref{oscillatorlength}).
%
%
%
Observe that we denoted energy as $E(\alpha)$ drawing attention to the 
fact that the value of this quantity depends on the parameter $\alpha$ 
as well as the spatial arrangement of the particle positions. 
%
%
%
%
%
%
%
It is easy to verify that $2 \, \pi/\lambda^2 = \alpha/l_{0}^2 $.
%
%

%
%
%
%
%
\begin{figure}[!thbp]
\begin{center}
\includegraphics[width=9cm]{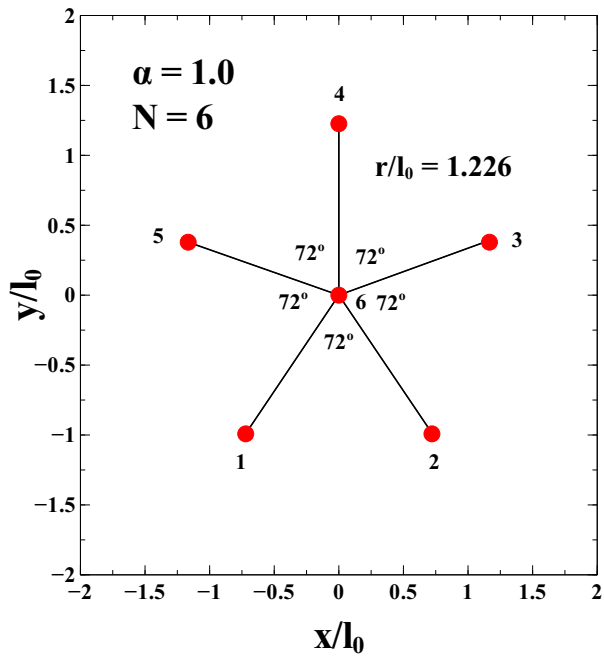}
\caption{                     Equilibrium configuration for a system of $N=6$ particles under 
                                  2D harmonic confinement for a value of 
                                  $\alpha={k_B T}/{(\hbar \, \omega)}=1.0$. 
                                  Dimensionless distances are used.
            }
\label{fign6}
\end{center}
\end{figure}

\section{Few-body systems}
\label{sec-fewbody}

Earlier results obtained using the Monte Carlo simulated annealing method 
for finite 2D systems of harmonically confined particles that interact with 
a statistical potential~\cite{aop} show complete agreement to the results
reported in Ref.~\cite{gajda1,gajda2} for systems  with $N=6$ and $N=3$ particles.
The most stable configuration for $N=6$ particles was a 
structure with one particle at the center of the harmonic trap 
and the other five particles forming a pentagon.
This arrangement agrees with that found in Ref.~\cite{gajda1}
where the shell radius of the pertinent pentagon structure 
was reported to be $r/l_{0}=1.265$.
The most stable configuration for $N=3$ particles was a crystalline
structure consisting of an equilateral triangle with its center
corresponding to the center of the harmonic trap.

Given the availability of quantitative results for $N=6$ particles,
we choose a system of $N=6$ particles as our principal 
case study to verify analytically the suitability of the approach.
The geometry of the $N=6$ crystalline system under consideration is
shown in Fig.~\ref{fign6}.
One can express all the distances of interest in terms of
%
\begin{equation} 
r=r_{1}=r_{2}=r_{3}=r_{4}=r_{5}  \ ,
\label{length-r}
\end{equation}
which represents the distance of each of the five particles 
located at the vertices of the pentagon relative to the center of the harmonic trap.
The sixth particle is located at the center of the harmonic trap. 
It simple to verify that:
%
%
%
%
%
%
%
%
\begin{equation} 
    \left\{
\begin{array}{ll}
r_{16}=r_{26}=r_{36}=r_{46}=r_{56}=r                             \ , \\ \\ 
r_{12}=r_{23}=r_{34}=r_{45}=r_{51}= 2 \, r \, \sin(36^{0})  \ ,  \\ \\
r_{13}=r_{24}=r_{35}=r_{41}=r_{52}= 2 \, r \, \sin(72^{0})  \ .
\end{array}
\right.    
\label{length-all}
\end{equation}

One can write the energy of the system in dimensionless units as:
\begin{widetext}
\begin{equation} \label{Etotal}
\frac{E (\alpha)}{\hbar \, \omega}=
- 5 \, \alpha \, \left\{  \ln \Bigl[ 1-\exp \Bigl( -\alpha \, \frac{r_{16}^2}{l_o^2}  \Bigr) \Bigr]+
                             \ln \Bigl[ 1-\exp \Bigl( -\alpha \, \frac{r_{12}^2}{l_o^2}  \Bigr) \Bigr]+
                             \ln \Bigl[ 1-\exp \Bigl( -\alpha \, \frac{r_{13}^2}{l_o^2}  \Bigr) \Bigr]
                     \right\}
+\frac{5}{2} \frac{r^2}{l_o^2} \ ,
\end{equation}
\end{widetext}
where $r_{16}$, $r_{12}$ and $r_{13}$ depend on $r$ 
and are given from Eq.(\ref{length-all}).
Since there are previously reported results for $\alpha=1$ we choose $\alpha=1$ 
and minimize the corresponding value of the energy,
$E(\alpha=1)/(\hbar \, \omega)$ with respect to $r$. 
%
The optimal value for $r$ found is:
\begin{equation} 
\frac{r}{l_{0}}=1.226 \ \ \ ; \ \ \ N=6 \ \ \ ; \ \ \ \alpha=1 \ .
\label{rfinal}
\end{equation}
This result compares favorably to the previously reported value of $r/l_{0}=1.265$
in single-shot measurements~\cite{gajda1}.
The relative percentage discrepancy is approximately $3 \,\%$.
The observed good quantitative agreement between the results 
is very rewarding considering 
the simplicity of the semi-classical model under consideration.

\begin{figure}[!thbp]
\begin{center}
\includegraphics[width=9cm]{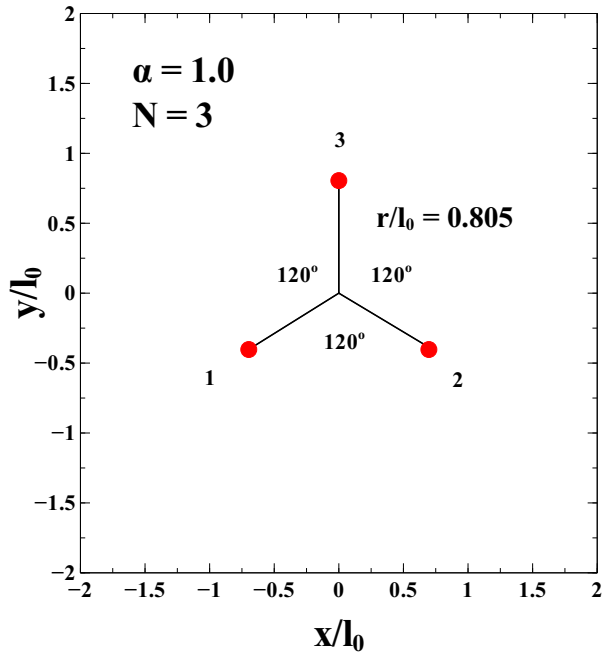}
\caption{                     Equilibrium configuration for a system of $N=3$ particles under 
                                  2D harmonic confinement for a value of
                                  $\alpha={k_B T}/{(\hbar \, \omega)}=1.0$. 
                                  Dimensionless distances are used.
             }
\label{fign3}
\end{center}
\end{figure}
The case of the equilateral triangle configuration of $N=3$ particles is simpler.
One can calculate the distance of each of the particles from the center of the trap 
analytically at any arbitrary value of $\alpha$ with the final result that reads:
\begin{equation} 
\frac{r}{l_{0}}=\sqrt{\frac{\ln \left( 6 \, \alpha^2+1\right)}{3 \, \alpha}} \ \ \ ; \ \ \ N=3 \ .
\label{rfinal3}
\end{equation}
As shown in Fig.~\ref{fign3},
the optimal radial distance from the center of the trap was found to be 
%
\begin{equation} 
\frac{r}{l_{0}}=0.805 \ \ \ ; \ \ \ N=3 \ \ \ ; \ \ \ \alpha=1 \ .
\label{rfinal-3}
\end{equation}
for the case of the system with $N=3$ particles at $\alpha=1$.
%
%
%
%
%
%
%
%

%
%
So far, the number of particles considered ($N=3$ and $6$ particles) 
is consistent with a closed shell numbering of the quantum system, 
i.e., the quantum many-body ground state is not degenerate.
It is also interesting to consider other numbers of particles, 
for example $N=4$ and $5$ particles
that correspond to an open shell structure where the quantum ground
state is degenerate.
The calculations for $N=4$ particles using the statistical interaction model, 
though not easy, allow for an analytical treatment.
We found that the lowest energy configuration for a system of $N=4$ 
harmonically confined particles interacting with a statistical potential
at a given value of parameter $\alpha$  
is that of a square with the particles located at the corners of the square 
and the center of the square coinciding with that of the trap.
The distance of each of particle from the center of the trap 
is suitably expressed in terms of the following parameter:
\begin{equation} 
z=\exp \Biggl[ -2 \, \alpha \, \left( \frac{r}{l_{0}} \right)^2 \Biggr]  \ \ \ ; \ \ \ N=4 \ ,
\label{rfinal4}
\end{equation}
where the value of $z$ that minimizes the energy is found to be:
\begin{equation} 
z= \frac{-2\ \alpha^2+\sqrt{1+8 \, \alpha^2+4 \, \alpha^4}}{1+8 \, \alpha^2} \ .
\label{rfinal4-2}
\end{equation}
The value of $z$ above represents the physically acceptable solution 
(the positive root) of a quadratic equation in terms of $z$
that is obtained during the minimization process of the energy.

An analytical treatment does not seem possible for $N=5$ particles.
However, for such a case, we found numerically that the minimum energy configuration
of the system is that of a pentagon where the center of the pentagon corresponds
to the center of the trap.
While there are no readily available quantitative results  
to whom we can compare our findings, the authors of Ref.[\citenum{gajda1}]
mention briefly that their single-shot imaging approach indicates that 
there are two equivalent configurations maximizing the
$5$-particle probability for the case of $N=5$ fermions, i.e. 
the case of an open shell structure.
For an open shell structure (case of $N=5$ particles), one has the freedom to
choose two occupied orbitals out of three quantum states leaving one empty.
To lift the degeneracy of the ground state, the authors of Ref.[\citenum{gajda1}]
assumed that the harmonic frequencies, $\omega_x$ and $\omega_y$ are slightly different.
The quantum energies in this case are
$E_{n_x \, n_y}=\hbar \, \omega_x \, (n_x+1/2)+\hbar \, \omega_y \, (n_y+1/2)$.
One can verify that 
$E_{n_x=0 \, n_y=2}-E_{n_x=2 \, n_y=0}= 2 \, \hbar \, ( \omega_y-\omega_x)$.
This means that the degeneracy of the ground state is exactly 
lifted if one assumes that, let's say $\omega_x$ is a little bit smaller than $\omega_y$,
namely, $\omega_y/\omega_x=1+\epsilon$ for a positive $\epsilon$.
This is the choice assumed in Ref.[\citenum{gajda1}] for $N=5$ particles
and, for such a choice, 
the empty orbital (the one with higher energy) is $n_x=0$ and $n_y=2$. 
For such a case, the two quantum crystalline configurations observed were
isosceles trapezoids differing by the reflection.
For such a choice, the quantum ground state has no rotational symmetry.
The only symmetry is reflection $y \rightarrow -y$.
The statistical interaction potential of our model in Eq.(\ref{Etotal}) 
leads to a minimum 
energy state with a circular shell structure of a pentagon for $N=5$ particles.
%
%
This means that in the case of $N=5$ particles, the quantum configuration
studied in Ref.[\citenum{gajda1}] is not appropriately captured by the 
statistical interaction.

Similar arguments as above can be applied to the case of $N=4$ particles.
Obviously, one should be aware that the statistical interaction potential
represents an approximation (at pair level) of the exact high-order quantum
correlations between quantum particles and, thus, has disadvantages and limitations.
While the main advantage of the method relies on its simplicity,
the reader should be alerted to the limitations of using a classical 
approach to a quantum problem.  
For example, based on the results for $N=5$ and $N=6$ particles,
it is probable that the description of the system in terms of the statistical
interaction is limited to particular cases of very symmetric confinements and 
very symmetric quantum states.
Alternatively, one also might argue
that the choice of the initial quantum configuration in Ref.[\citenum{gajda1}]
may have influenced what sort of Pauli crystalline structure is observed at the end.
%
%


%
%
%
%
\begin{figure}[!thbp]
\begin{center}
\includegraphics[width=9cm]{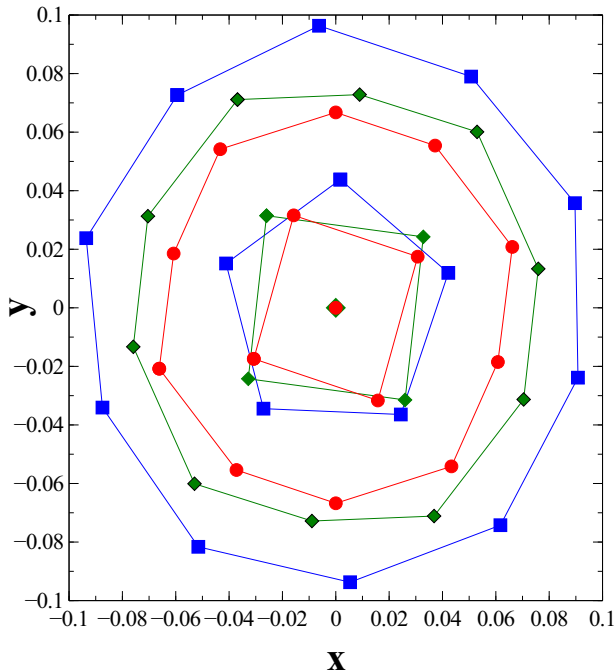}
\caption{      Plot of the minimum energy configuration for $N=15$ particles 
                  under 2D harmonic confinement
                  for different values of parameter $\alpha={k_B T}/{(\hbar \, \omega)}$. 
                  For $\alpha=1.5$ (circles, red lines) the particles form a 
                 shell structure with a $(1,4,10)$ configuration. 
                 For $\alpha=2.0$ (diamonds, green lines) 
                 we still have the same $(1,4,10)$ configuration as before. 
                 However, for $\alpha=3.0$ (squares, blue lines) the
                 configuration of the particles changes to a $(5,10)$ one.  
                 Distances are measured in arbitrary units.
            }
\label{fign15}
\end{center}
\end{figure}
\begin{figure}[!thbp]
\begin{center}
\includegraphics[width=9cm]{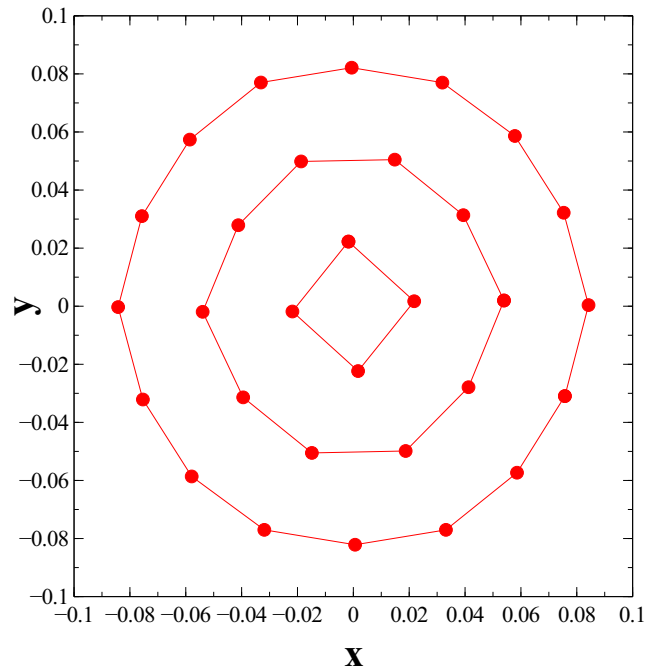}
\caption{                      Plot of the minimum energy configuration 
                                  for a system of $N=30$ particles under 
                                  2D harmonic confinement for a value of 
                                  $\alpha={k_B T}/{(\hbar \, \omega)}=1.0$. 
                                  Distances are measured in arbitrary units.
            }
\label{fign30}
\end{center}
\end{figure}
%
%
%
%
The reason why we studied small values of $N$ in this work
is to generate results that can be analytically checked up to a good degree.
Obviously, for larger systems of particles, it is not possible
to find  the minimum energy configuration by using analytical tools.
For such a scenario, the entire approach is considerably 
more difficult and should be fully numerical.
In order to get glimpses of the behavior of the system at larger values of $N$,
we carried out additional numerical calculations using the 
simulated annealing method~\cite{annealing} 
in order to find the minimum energy configuration for some of the largest 
values of $N$ that we could reliably simulate.
%

The minimum energy configurations 
for a system of $N=15$ harmonically confined particles
interacting with a statistical potential at
various values of the parameter 
$\alpha={k_B T}/{(\hbar \, \omega)}=1.5, 2.0$ and $3.0$
are shown in Fig.~\ref{fign15}.
The shell structure $(1,4,10)$ is found for $\alpha=1.5$. 
Such a shell structure is preserved as we increase the temperature 
to a corresponding value of $\alpha=2.0$.
Note that this shell structure is not the same as that of 
the same number of charged particles which are confined 
by a circular parabolic well and interact 
with a Coulomb potential~\cite{bedanov1994}.
Thus, it is apparent that at low temperatures (where the statistical
interaction potential is more effective on capturing part of the quantum 
correlations originating from the Fermi statistics) 
the most stable configuration of the system
has a different structure from that of a Wigner crystal of charged 
particles confined by a circular parabolic potential.
However, this does not mean that the statistical potential will always lead to the same 
configuration or is equivalent to the structures observed in Pauli crystals.
In fact, the $(1,4,10)$ geometry of the configuration obtained for $N=15$ particles
is also different from the $(1,5,9)$ shell structure 
of the corresponding Pauli crystal reported in Ref.[~\citenum{rakshit,gajda1}].  
As stated earlier, the statistical interaction potential is a classical approximation
at the two-body level of the high order correlations originating from the Fermi statistics.
Therefore, one should neither expect, nor suggest that the statistical potential
can reproduce the high order correlations originating from quantum Fermi statistics.   
As we increase the temperature further to a larger value of $\alpha=3.0$
the more stable configuration structure changes to $(5,10)$
that would be that of a 2D system of charged classical particles with 
a parabolic confinement \cite{bedanov1994}, namely, 
a classical model of a Wigner crystal in a parabolic confinement potential.  
A larger value of $\alpha$ implies that the statistical potential is less relevant
at typical separation distances between particles and, thus, 
the overall structure of the 
configuration tends to resemble that of repelling classical particles 
under a parabolic confinement potential. 
%
%
%

In Fig.~\ref{fign30} we show the minimum energy configuration 
for the largest system that we were able to simulate.
This system consists of $N=30$ harmonically confined particles
interacting with a statistical interaction potential for $\alpha=1.0$.
We noticed that the stabilized shell structure, $(4, 10, 16)$
differs from the Wigner crystal one~\cite{bedanov1994} 
when no statistics is involved, that is, $(5, 10, 15)$.  
This result suggests that the most stable configurations 
for high number of particles (for instance, $N=30$ particles)
at sufficiently low temperatures 
is not that of a Wigner crystal of parabolically confined charges \cite{bedanov1994}.
However, a structure that has a geometry different from that of the 
Wigner crystal configuration at $N=30$ particles may be 
or may not be the same as the Pauli crystal counterpart (as seen earlier for the case of $N=15$ particles).  
We are unsure if this the situation for the case of $N=30$ particles since there are no available 
quantum mechanical single-shot data results to whom we can compare.
In a nutshell, despite the fact that our approach is classical and has its limitations,
some interesting insights do come from the numerical results that we obtained
for systems with $N=15$ and $N=30$ particles.
%
%
%


\section{Discussion and conclusions}
\label{sec-discussion}

\begin{figure}[!thbp]
\begin{center}
  \includegraphics[width=6cm]{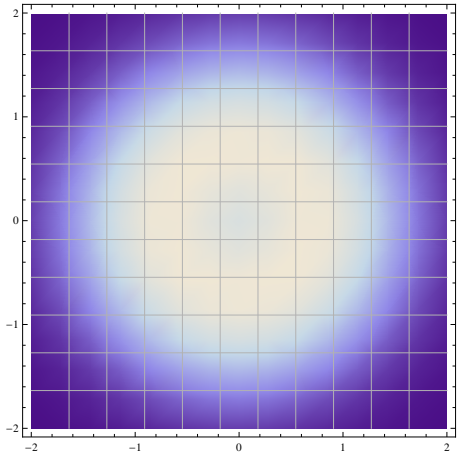}
\caption{            One-particle density function for a system of $N=3$ fermions 
                        at their ground state. 
                        Brighter colors imply larger values of the function. }
\label{fig-den}
\end{center}
\end{figure}

From the theoretical point of view, a good case for the existence of Pauli crystals 
should be based on showing that these structures provide an appropriate 
description of relevant aspects of a full quantum mechanical solution of the
few-body problem under consideration.
At zero temperature, the quantum state of a non-interacting system of $N$ confined
fermions is given by the Slater determinant wave function in Eq.(~\ref{slater}).
At a finite temperature, the quantum description of the system is provided
by a thermal state, corresponding to the canonical density matrix.
Does any of these two quantum solutions (for instance, the ground state
for a small system of $N=3$ fermions at $T=0$) resemble a Pauli crystal?
The answer is not affirmative if one looks at the form of the one-particle density function,
$\rho(x,y)=|\phi_{00}(x,y)|^2+|\phi_{10}(x,y)|^2+|\phi_{01}(x,y)|^2$ for $N=3$
which is plotted in Fig.~\ref{fig-den}.
However, it has been already noted in Ref.[\citenum{rakshit,gajda1,gajda2}] 
that the one-particle density function
cannot reveal any underlying geometric arrangement of the particles. 
Instead, one should look at the (conditional) $N$-particle
correlation function if there are any underlying geometric arrangements
(for instance, see Figure 2 of Ref.~[\citenum{rakshit}]).
Based on these explanations, the answer to the question 
of whether such a geometric arrangement of fermions due to quantum 
statistics is firmly based on a theoretical model is still lacking.

With that said, we take the opportunity to iterate again that the objective 
of this work is simpler than a full-fledged theoretical investigation
of the possible existence (or not) of the Pauli crystals.
The focal point of this work was a practical one and had as its main objective 
to consider and investigate a simple model that may be able to capture 
both qualitatively and quantitatively 
all the elements of the physics of few-body Pauli crystals under the same 
conditions as recently reported in the literature~\cite{rakshit,gajda1,gajda2}.
%
%
The basic idea of the approach is the mapping of non-interacting fermions 
into a classical interacting system via the so-called 
effective statistical interaction potential method~\cite{uhlenbeck}.
Such an effective interaction between the classical particles 
mimics the Pauli repulsion effects.
We considered few-body systems of $N=3$ and $N=6$ particles  
under 2D harmonic confinement as well as some larger systems.
%
%
The most energetically stable structure for $N=6$ particles 
and the equilibrium shell radius of the pentagon structure for $N=6$ particles
calculated at a temperature that corresponds to $\alpha=1$ 
was found to compare very favorably to the reported value
obtained from single-shot experiments~\cite{gajda1}.
%
%
%
%
%
%
%
The current results for $N=3$ and $N=6$ particles 
are not in disagreement with the idea of Pauli crystals.
However, results for open shell structures ($N=4$ and $N=5$ particles) suggest
that the question of whether such crystalline structures do really exist in a quantum
system or whether they are artifacts of the configuration density
approach used is legitimate question open to discussion.

The nature of the statistical interaction potential adopted in this work 
is such that it is relevant for interparticle distances $r_{ij}<\lambda \propto 1/\sqrt{T}$.
Therefore, an experimental observation of Pauli crystals would be possible only 
for very low temperatures in systems of non-interacting fermions
such as confined charge-neutral fermionic atoms. 
Such systems are known that can be cooled to very low temperatures.
By the same token, confined electronic states that can be created in
semiconductor quantum dots~\cite{scripta} or 
oxide quantum wells~\cite{stemmer} 
would not be good choices since, most likely, 
the Coulomb interaction effects between electrons 
will domininate over quantum statistics at any given temperature.
%
%
%
%
%
The effective statistical potential involves a temperature-dependent 
characteristic length scale known as the mean thermal wavelength, 
$\lambda$ given in Eq.(\ref{lambdaT}).
At $T=0$, the mean thermal wavelength becomes infinite.
This means that we cannot reliably extend the use of the statistical potential
to temperatures in the $T \rightarrow 0$ limit. 
The statistical potential has a thermodynamic origin and, thus, 
depends on the consideration of a nonzero temperature.
Therefore, the reader should be aware of the limitations of the method.
After all, the statistical interaction model represents
a classical approximation to the real high-order quantum correlations 
between particles since it includes only pair-wise correlations. 

Note that the general expression for the energy of $N=3$ and $N=6$ particles 
is initially given for an arbitrary $\alpha$ and then numerically minimized for $\alpha=1$.
On the other hand,
numerical simulations for larger systems with $N=15$ and $N=30$ particles 
involve different values of $\alpha$, respectively, $\alpha=1.5, 2.0, 3.0$ ($N=15$)
and $\alpha=1$ (the $N=30$ case). 
This means that, in our calculations, we considered different choices 
of the temperature parameter $\alpha$ with no particular focus 
ony specific value.
The choice $\alpha=1$ encountered in few places of our study such as 
in Eq.(\ref{rfinal}) or Eq.(\ref{rfinal-3}) is because 
of the availability of  results for $\alpha=1$ in Ref.[\citenum{rakshit,gajda1,gajda2}].
For example, we can immediately compare our analytical result for $\alpha=1$
in Eq.(\ref{rfinal}), namely $r/l_0=1.226$ for $N=6$ to the corresponding
value of $r/l_0=1.265$ at $\alpha=1$ as reported in Ref.[\citenum{gajda1}]. 
In this sense, there no deep physical argument for the choice of $\alpha=1$
except the opportunity for a direct comparison of 
the results of this work to corresponding results in the literature
(reported at $\alpha=1$).
Overall, the results suggest that the use of this model has advantages 
due to its simplicity, but also it has limitations.
It is probable that an accurate description of the system in terms of the statistical
interaction may be limited to particular cases of symmetric quantum states.
Furthermore, being a classical approximation at two-body level, 
the statistical interaction potential may not be able to fully reproduce 
all the exact high-order correlations between quantum particles
arising from the Fermi statistics.

\vspace{0.5cm}

\section*{Acknowledgements}

The research of 
one of the authors (O. C.)
was supported in part by National Science Foundation (NSF) grant no. DMR-1705084.



\section*{Conflict of interest}
The authors declare no conflict of interest.



\begin{thebibliography}{999}


\bibitem{pauli}
W. Pauli,
{\it Z. Physik}  \textbf{1925},  {\it 31}, 765. 



%
%
%
%
\bibitem{dirac}
P. A. M. Dirac,
{\it Proc. R. Soc. Lond. A}  \textbf{1929}, {\it 123}, 714.

 
\bibitem{contemporary}
D. D. Vvedensky, S. Crampin, M. E. Eberhart, J. M. Maclaren,
{\it Contemp. Phys.} \textbf{1990}, {\it 31}, 73. 
%
%
%
%




\bibitem{cohen}
M. L. Cohen,
{\it MRS Bull.} \textbf{2015},  {\it  40}, 516. 
%
%

%


\bibitem{2014c}
O. Ciftja,
{\it J. Phys. Chem. Sol.} \textbf{2014}, {\it 75}, 931. 




%



\bibitem{slater}
J. C. Slater, 
{\it Phys. Rev.} \textbf{1929},  {\it  34}, 1293. 




\bibitem{2015f}
O. Ciftja,
{\it AIP Adv.} \textbf{2015}, {\it 5}, 017148. 

%



\bibitem{rakshit}
D. Rakshit, J. Mostowski, T. T. Sowi\'nski, M. Zaluska-Kotur,  M. Gajda,
{\it Sci. Rep.} \textbf{2017},  {\it 7}, 15004. 


\bibitem{gajda1} 
M. Gajda, J. Mostowski, T. Sowi\'nski, M. Zaluska-Kotur,
{\it Europhys. Lett.} \textbf{2016},  {\it 115}, 20012. 



\bibitem{gajda2} 
M. Gajda, J. Mostowski, T. Sowi\'nski, M. Zaluska-Kotur, 
{\it arXiv:1511.01036v3}  \textbf{2015}.


\bibitem{Wig} 
E. P. Wigner, 
{\it Phys. Rev. B} \textbf{1934}, {\it 46}, 1002. 


\bibitem{Grimes} 
C. C. Grimes, G. Adams, 
{\it Phys. Rev. Lett.}  \textbf{1979}, {\it 42}, 795. 



%


\bibitem{uhlenbeck} 
G. E. Uhlenbeck, L. Gropper, 
{\it Phys. Rev.} \textbf{1932},  {\it 41}, 79. 


\bibitem{aop}
J. Batle, O. Ciftja, A. Farouk, M. Alkhambashi, S. Abdalla, 
{\it Ann. Phys.} \textbf{2017}, {\it 384}, 11. 

%




\bibitem{annealing}
S. Kirkpatrick, C. D. Gelatt,  M. P. Vecchi,
{\it Science} \textbf{1983}, {\it 220}, 671. 



\bibitem{bedanov1994}
V. N. Bedanov, F. M. Peeters,
{\it Phys. Rev. B} \textbf{1994},  {\it 49}, 2667. 









%
%

\bibitem{scripta}
O. Ciftja, 
{\it Phys. Scr.}   \textbf{2013}, {\it 88}, 058302. 

%

\bibitem{stemmer}
S. Stemmer, A. J. Millis,
{\it MRS Bull.} \textbf{2013}, {\it 38}, 1032. 
%
%
%


\end{thebibliography}
\end{document}